\documentstyle[sprocl]{article}

\input{psfig}
\def\Journal#1#2#3#4{{#1} {\bf #2}, #3 (#4)}

\def\NPB{{\em Nucl. Phys.} B}
\def\PLB{{\em Phys. Lett.}  B}
\def\PRL{\em Phys. Rev. Lett.}
\def\PRD{{\em Phys. Rev.} D}
\def\ZPC{{\em Z. Phys.} C}

\def\be{\begin{equation}}
\def\ee{\end{equation}}
\def\bea{\begin{eqnarray}}
\def\eea{\end{eqnarray}}
\def\gev{\rm GeV}
\def\tev{\rm TeV}

\def\epem{e^+e^-}
\def\emem{e^-e^-}

\def\zl{Z_L^{}}
\def\wlwl{W_L^{}W_L^{}}
\def\wpwm{W^+W^-}
\def\tr{{\rm Tr}}

\begin{document}
\rightline{MADPH-99-1141}
\rightline{hep-ph/9910495}
\vskip 0.2in

\title{STRONG INTERACTIONS OF WEAK BOSONS
 \footnote{Plenary talk presented at the 5$^{th}$ 
International Linear Collider 
Workshop, Sitges, Barcelona, Spain, May 1999.}
}

\author{T. HAN}

\address{Department of Physics, University of Wisconsin,\\ 
1150 University Avenue, Madison, WI 53706, USA}


\maketitle\abstracts{We discuss the parameterization for
electroweak gauge boson interactions without a light Higgs
boson. We present the constraints on the anomalous 
gauge-boson
couplings from the current experiments. We emphasize that
the four-point couplings involving the longitudinal weak
bosons are genuine to the underlying strong dynamics 
responsible for the electroweak symmetry breaking. We
study the sensitivity to the four-point couplings and
the possible heavy resonant states in this sector
at future TeV $\epem$ linear colliders.
}
  
\section{Introduction}

The most prominent mystery in contemporary particle physics
is the origin of electroweak symmetry breaking (EWSB) and the mass
generation mechanism for fermions. The celebrated Standard
model (SM), which has passed the experimental test with high 
precision up to the energy scale 
${\cal O}(100\ \gev)$, fulfills the job by 
introducing an effective potential
\be
V(\Phi) = \lambda ( |\Phi|^2-v^2/2 )^2.
\ee
The scalar Higgs doublet can be parameterized by
\[ \Phi = {\rm exp}(iw^a \tau^a/2v) \left( \begin{array}{c}
                  0 \\ (v+H)/\sqrt 2
                 \end{array} \right), \]
where $w^a\ (a=1,2,3)$ are the Goldstone bosons and $H$ the
physical Higgs boson of mass $m_H=\sqrt{2\lambda} v$. The vacuum 
expectation value (vev) $v\approx 246\ \gev$ sets the mass 
scale not only for the electroweak gauge bosons, but also
for all fermions through Yukawa interactions. Thus searching
for the Higgs boson has become the top priority to 
understand the electroweak symmetry breaking and mass 
generation mechanisms. However, due to the unknown parameter 
$\lambda$ in the Higgs potential, the Higgs boson mass $m_H$ 
is a free parameter. Current experimental searches at LEP2
have set a limit $m_H>98.8\ \gev$ at a 95\% 
confidence level.\cite{lepmh} Theoretically, the SM cannot
be a consistent effective theory if $m_H> 800\ \gev$. Thus,
if there is no light Higgs boson found in the collider
experiments, then new strong dynamics must set in. It has
been pursued actively to explore the possible new dynamics
responsible for the EWSB both theoretically and experimentally.

Although the original idea of Technicolor \cite{tc}
for dynamical electroweak symmetry breaking is attractive, 
one would need the Extended Technicolor \cite{etc} to
give fermions their masses. Tremendous efforts have been 
made to incorporate the vast fermion mass hierarchy yet to 
avoid severe flavor changing neutral currents.\cite{walk}
Precision electroweak measurements at the $Z$ pole
can also impose significant constraint on the new
strong dynamics, such as on the number of Technicolors
and the SU(2) breaking effects.\cite{stu} 
The fact that the top-quark mass is
miraculously close to the electroweak scale makes it very
attempting to consider the role of the top quark in the
EWSB,\cite{bhl} and the new interaction of topcolor has
been also introduced to account for the EWSB and the 
top-quark mass generation.\cite{topc,tc2}
More recently, models of dynamical symmetry
breaking with seesaw mechanism of quark condensation are
proposed.\cite{seesaw} Models of dynamical symmetry
breaking often lead to predictions of rich phenomenology.
Technicolor theories generically result in technihadrons like
$\pi_T, \eta_T, \rho_T, A_{1T}$ and $\omega_T$. The
topcolor models often have colorons ($Z',\ V_8$); while
the top seesaw models have flavorons ($\chi^{}_{L,R},\ F's$)
and other composite scalars in the spectrum. If these
particles exist well below 1 TeV, the experiments at the
next generation of colliders will be able to discover them
by their distinctive production and decay processes.

In this talk, we would like to take a different direction,
namely a relatively model-independent approach to parameterize
the Strongly-interacting Electro-Weak Sector (SEWS). One thing
we know for sure in the gauge boson sector is that the 
longitudinally polarized states
($W_L^\pm,\zl$) exist and they possess an (approximate) SU(2)
custodial symmetry, leading to the mass relation
$M_W\approx M_Z\cos\theta_W$ where $\theta_W$ is the
weak mixing angle. In the scenario of 
spontaneous EWSB, the states $W_L^\pm,\zl$ are
equivalent to the Goldstone bosons $\omega^\pm,\omega^0$
at high energies.\cite{sews,eqv} One can thus construct an
electroweak effective chiral Lagrangian based on the
Goldstone bosons to parameterize the EWSB sector.
One can also consider to include higher dimensional
operators by the derivative expansion, which would lead
to the ``anomalous couplings'' beyond the restricted 
form of the SM for the gauge bosons.
We can also go a step further to introduce lower-lying 
heavy resonances near the TeV region. These will be
discussed in the next section. In Sec.~\ref{sec:current},
we summarize the constraints on the anomalous coupling
parameters from the current experimental results. We
present analyses for the sensitivity to the SEWS sector
at future high energy/high luminosity $\epem$ colliders
in Sec.~\ref{sec:quartic}, where we emphasize the important
role for the processes $\wlwl \to \wlwl$ as well as
$\wlwl\to t\bar t$. We make some
concluding remarks in Sec.~\ref{sec:conclude}.
For more discussions, there are recent reviews \cite{review} 
that dealt with related topics.

\section{Model-independent parameterization for SEWS}\label{sec:nomodel}

It is known that the most economical description of EWSB
below a new physics scale is the electroweak chiral
Lagrangian with non-linear realization of the symmetry.\cite{ew} 
The lowest order term respecting the symmetry can be written as
\be
{\cal L}^{(2)}={v^2\over 4} \tr[(D^\mu U)^\dagger (D_\mu U)],
\ee
where
\be
U=\exp[i\tau^a\omega^a/v],\qquad 
D_\mu U=\partial_\mu U + igW_\mu^a{\tau^a\over 2} U - 
ig' U B_\mu {\tau^3\over 2}.
\ee
${\cal L}^{(2)}$ breaks the electroweak gauge symmetry 
spontaneously and the coefficient
is fixed by the gauge-boson mass.
It also respects the custodial SU(2) symmetry 
and the low-energy theorem \cite{let} if gauge coupling $g'$
is ignored. To gain more information of the strong dynamics
responsible for the EWSB, we need to go beyond the minimal
term of ${\cal L}^{(2)}$. 
It should be mentioned that there is yet another popular
approach to the electroweak effective Lagrangian, namely 
the linear realization with an explicit doublet Higgs 
field.\cite{linear} For the sake of simplicity, we will 
not discuss this approach. In the rest of this 
section, we will focus on some parameterization for the couplings
among the Goldstone bosons, some lower-lying resonances
as well as the top quark.

\subsection{Scalar resonance}

One can parameterize an isosinglet heavy scalar resonance 
and its coupling to the Goldstone bosons and the top quark as
\be
{\cal L}^H = {1\over 2}\partial^\mu H\partial_\mu H - 
 {1\over 2}{M_H^2} H^2 + {1\over 2}
(g_h^{} v H+g_h'{H^2\over 2}) \tr \partial^\mu U^\dagger \partial_\mu U
 -{\kappa_t} {m_t\over v} H\bar t t.
\ee
There are four free parameters in this framework 
$M_H,\ g_h,\ g_h'$ and $\kappa_t$. The couplings $g_h$ and $\kappa_t$
can be traded to the partial decay widths to $\wlwl$ and $t\bar t$,
characterizing the dynamics of the underlying theory.
This Lagrangian reduces to the SM Higgs sector when 
$g_h=g_h'=\kappa_t=1$. 

\subsection{Vector resonance}

There is a systematic way to introduce an isotriplet vector 
resonance into the chiral Lagrangian,\cite{techrho} with an example 
called BESS \cite{bess} for the EWSB sector.
There are three free parameters that can be expressed 
by the decay partial widths to $\wlwl$ and to fermions, plus
the vector boson mass $M_V$.\cite{bess,baggeretal} 
We note that other lower-lying vector 
states such as $A_1$ and $\omega_T$ can also be incorporated 
in the chiral Lagrangian.\cite{a1,hhh,egamma}

\subsection{Next-to-leading order terms}

If the resonant states are kinematically unaccessible, then
the low-energy effects can be conveniently parameterized by
the coefficients of higher order terms of the electroweak
chiral Lagrangian as the derivative expansion.
There are 12 $SU_L(2)\otimes U_Y(1)$ gauge
invariant and CP-even operators,\cite{ew} parameterized by 
\bea
{\cal L}^{(2)\prime} & =& \ell_0 (\frac{v}{\Lambda})^2~\frac{v^2}{4}
               [ {\rm Tr}({\cal T}{\cal V}_{\mu})]^2 ~~,\nonumber\\
{\cal L}_1 & = &\ell_1 (\frac{v}{\Lambda})^2~ \frac{gg^\prime}{2}
B_{\mu\nu} {\rm Tr}({\cal T}{\bf W^{\mu\nu}}) ~~,\nonumber\\
{\cal L}_2 & =& \ell_2 (\frac{v}{\Lambda})^2 ~\frac{ig^{\prime}}{2}
B_{\mu\nu} {\rm Tr}({\cal T}[{\cal V}^\mu,{\cal V}^\nu ]) ~~,\nonumber\\
{\cal L}_3 & =& \ell_3 (\frac{v}{\Lambda})^2 ~ig
{\rm Tr}({\bf W}_{\mu\nu}[{\cal V}^\mu,{\cal V}^{\nu} ]) ~~,\nonumber\\
{\cal L}_4 & =& \ell_4 (\frac{v}{\Lambda})^2 
                     [{\rm Tr}({\cal V}_{\mu}{\cal V}_\nu )]^2 ~~,\nonumber\\
{\cal L}_5 & =& \ell_5 (\frac{v}{\Lambda})^2 
                     [{\rm Tr}({\cal V}_{\mu}{\cal V}^\mu )]^2 ~~,\nonumber\\
\label{ls}
{\cal L}_6 & =& \ell_6 (\frac{v}{\Lambda})^2 
[{\rm Tr}({\cal V}_{\mu}{\cal V}_\nu )]
{\rm Tr}({\cal T}{\cal V}^\mu){\rm Tr}({\cal T}{\cal V}^\nu) ~~,\\
{\cal L}_7 & =& \ell_7 (\frac{v}{\Lambda})^2 
[{\rm Tr}({\cal V}_\mu{\cal V}^\mu )]
{\rm Tr}({\cal T}{\cal V}_\nu){\rm Tr}({\cal T}{\cal V}^\nu) ~~,\nonumber\\
{\cal L}_8 & =& \ell_8 (\frac{v}{\Lambda})^2~\frac{g^2}{4} 
[{\rm Tr}({\cal T}{\bf W}_{\mu\nu} )]^2  ~~,\nonumber\\
{\cal L}_9 & =& \ell_9 (\frac{v}{\Lambda})^2 ~\frac{ig}{2}
{\rm Tr}({\cal T}{\bf W}_{\mu\nu}){\rm Tr}
        ({\cal T}[{\cal V}^\mu,{\cal V}^\nu ]) ~~,\nonumber\\
{\cal L}_{10} & =& \ell_{10} (\frac{v}{\Lambda})^2\frac{1}{2}
[{\rm Tr}({\cal T}{\cal V}^\mu){\rm Tr}({\cal T}{\cal V}^{\nu})]^2 ~~,
\nonumber\\
{\cal L}_{11} & =& \ell_{11} (\frac{v}{\Lambda})^2 
~g\epsilon^{\mu\nu\rho\lambda}
{\rm Tr}({\cal T}{\cal V}_{\mu}){\rm Tr}
({\cal V}_\nu {\bf W}_{\rho\lambda}) ~~, \nonumber
\eea
where 
${\bf W}_{\mu\nu} =\partial_{\mu}{\bf W}_{\nu}-\partial_{\nu}{\bf W}_{\mu}
                      +ig [{\bf W}_\mu , {\bf W}_\nu ]$, 
$B_{\mu\nu}= \partial_\mu B_\nu - \partial_\nu B_\mu$
and
${\cal V}_{\mu}\equiv (D_{\mu}U)U^\dagger~$,  
$~{\cal T}\equiv U\tau_3 U^{\dagger} ~$. The cutoff $\Lambda$
characterizes the scale for the effective theory at which
the new underlying dynamics sets in, presumably around 
$\Lambda={\rm min}(4\pi v,M_{H,V}^{})$. 
In such a normalization and without other
symmetries in the underlying physics, one would expect
the natural size for the couplings to be 
$\ell_i \sim {\cal O}(1)$.

\subsection{Remarks on the anomalous couplings}

The operators in Eq.~(\ref{ls}) modify the gauge boson interactions,
leading to the so-called ``anomalous couplings''. Conventionally,
the anomalous couplings of the gauge boson interactions are formulated
by a Lorentz and electromagnetic gauge-invariant effective 
Lagrangian \cite{hhpz}
\bea
{{\cal L}_V\over g_V} 
&=& i{g_1^V} \left( W_{\mu\nu}^\dagger W^\mu V^\nu 
      - W_\mu^\dagger V_\nu W^{\mu\nu} \right) 
+ i {\kappa_V} W_\mu^\dagger W_\nu V^{\mu \nu \nonumber}\\ 
&+& i {{\lambda_V} \over m_W^2} 
W_{\lambda \mu }^\dagger W^\mu{}_\nu V^{\nu \lambda }
- {g_4^V} W_\mu^\dagger W_\nu \left( \partial^\mu V^\nu +
\partial^\nu V^\mu \right) \\ 
& +& {g_5^V} \epsilon^{\mu\nu\lambda \rho} 
\left( W_\mu^\dagger \partial_\lambda W_\nu -
\partial_\lambda W_\mu^\dagger \, W_\nu 
\right) V_\rho 
 + i {\tilde \kappa_V} 
W_\mu^\dagger W_\nu \tilde V^{\mu \nu }\nonumber\\ 
&+& i {{\tilde \lambda_V} \over m_W^2} 
W_{\lambda \mu }^\dagger W^\mu{}_\nu \tilde V^{\nu \lambda } .
\nonumber
\eea 
In the SM at tree level,
$g_1^V = \kappa_V = 1, \lambda_V = \tilde 
\lambda_V = \tilde \kappa_V = g_4^V = g_5^V = 0.$
The deviation from the SM values can be expressed in terms of 
the coupling relations between this formalism and the electroweak
chiral Lagrangian as
\bea
&&
\nonumber
{\Delta g_1^Z}= {v^2\over \Lambda^2}
({\ell_0\over c_{2w}}+
{e^2{\ell_1}\over c^2_w c_{2w}}+
{e^2{\ell_3}\over c^2_w s^2_w}),\qquad 
g_5^Z= {v^2\over \Lambda^2} {e^2{\ell_{11}}\over c^2_w s^2_w},
\\ 
&&{\Delta \kappa_\gamma}={v^2\over\Lambda^2}{e^2\over s^2_w}
({-\ell_1+\ell_2+\ell_3-\ell_8+\ell_9}), \\ 
&& {\Delta \kappa_Z}= {v^2\over \Lambda^2}
({{\ell_0}\over c_{2w}}+
{2e^2{\ell_1}\over c_{2w}}-
{e^2{\ell_2}\over c^2_w} +
{e^2\over s^2_w}[{\ell_3-\ell_8+\ell_9}]),
\nonumber
\eea
where $s_w=\sin\theta_W^{},\ c_{2w}=\cos2\theta_W^{}$.

Note that these couplings are three-point interactions
and always involve some pure gauge fields that may not be
sensitive to the EWSB sector. This is especially true
for those couplings involving photons,
no matter it is a triple or quartic coupling. 
In contrast, the couplings
$\ell_{4,5,6,7,10}$ are of particular interests: They are
the {\em genuine} Goldstone boson interactions which 
characterize the underlying physics responsible for the
EWSB.\cite{ew,lsvalues}
Furthermore, those couplings are expected to be
enhanced \cite{ew,power} over the gauge couplings due to 
the new strong dynamics.\cite{vdb} For instance,\cite{ours}
\bea
&&{\rm heavy\ scalar}\ M_H=2\ \tev:\quad 
\ell_4\approx 0.0,\quad \ell_5\approx 0.33;\nonumber\\
&&{\rm heavy\ vector}\ M_V=2\ \tev:\quad
\ell_4\approx 0.38\quad \ell_5\approx -0.31.\nonumber
\eea

\section{Current constraints on SEWS}\label{sec:current}

Physics associated with vector boson pairs has been
experimentally studied both at the Tevatron \cite{tevatron}
and at LEP2.\cite{lepew} The triple gauge-boson couplings 
can be measured through those processes. With good agreements
with the SM expectation, the constraints from LEP2, 
CDF and D0 on the relevant anomalous 
couplings at a 2$\sigma$ level are \cite{lepew}
\[ \left\{ \begin{array}{c}
-0.07 <\Delta g_1^Z< 0.05\ \Longrightarrow -8.4 <\ell_3< 4.5\\ 
-0.11<\Delta \kappa_\gamma <0.23\ \Longrightarrow -16 <\ell_{2,9}< 35
          \end{array} \right.  \]
for $\Lambda=2\ \tev$. So far, the vector boson pairs produced 
are dominantly transversely polarized from the light fermion
radiation, independent of the Higgs sector. 
Although the constraints will be
further tightened up with more LEP2 data coming out, one
may only expect to discover new physics signal through those
channels when there is a resonance accessible or nearby.

Turning to the EWSB sector, besides the Higgs mass limit 
from the direct search,\cite{lepmh} 
precision electroweak data can be used to infer a 
Higgs mass limit $M_H<107^{+67}_{-45}$ GeV,\cite{mh} 
which is only valid for a SM-like Higgs boson.
Furthermore, from oblique corrections at the $Z$ pole, 
one can relate the $S,T,U$ parameters \cite{stu} to the
anomalous couplings at tree level
\be
S=-{1\over \pi}({4\pi v\over \Lambda})^2\ell_1,\qquad
T= {1\over 2\pi e^2}({4\pi v\over \Lambda})^2\ell_0,\qquad
U=-{1\over \pi}({4\pi v\over \Lambda})^2\ell_8,
\ee
and translate the limits on $S,T,U$ \cite{mh} to 
\[ \left\{ \begin{array}{c}
S= -0.27\pm 0.12\ \Longrightarrow  0.04<\ell_1<0.67\\ 
T= 0.00\pm 0.15\  \Longrightarrow -0.08<\ell_0<0.08 \\
U= 0.19\pm 0.21\  \Longrightarrow -0.80<\ell_8<0.30
          \end{array} \right.  \]
for $\Lambda=2\ \tev$ and at 2$\sigma$. We see that the 
constraint from $T$ on the custodial symmetry breaking 
effect is very strong.\footnote{A new analysis appeared
more recently in this topic.\cite{bfs}} 
Loop corrections to the $T$ 
parameter through other couplings have been evaluated.\cite{eboli}
With the new $T$ value above, the limits on the couplings
of interest are
\begin{eqnarray}
&&-11 <\ell_4< 11,\ \quad - 28<\ell_5< 28,\ \quad  -1.4<\ell_6<1.4,
\nonumber \\ 
&&-11<\ell_7<11,\ \quad -1.5<\ell_{10}<1.5.
\nonumber
\end{eqnarray}
This calculation is in principle similar to that
by considering the $Z$ partial widths.\cite{zwidth}

It is interesting to note that the couplings $\ell_{6,7,10}$
violate the custodial SU(2) symmetry, yet the current
constraints on them are not very strong. It would be of
great theoretical significance if one observed the
custodial SU(2) symmetry breaking in SEWS.
Currently, low energy constraints on heavy resonances are
still weak with one exception for a vector resonance like
$\rho_{T}^{}$ which may mix with $\gamma,Z$. However, if
there is also a nearly degenerate axial vector $A_1$,
then the constraint on $\rho_{T}^{}$ can be
avoided.\cite{a1}

\section{Quartic gauge-boson couplings at linear colliders}\label{sec:quartic}

Physics potential at $\epem$ colliders has been nicely presented
in recent review articles \cite{PMZ} and in many talks
in this workshop.
To explore the physics of the electroweak symmetry breaking sector, 
the {\em most direct} way is to study the four-gauge boson couplings 
to probe the genuine Goldstone boson interactions, no matter
there is a light Higgs boson or not. The relevant processes and 
the corresponding couplings of our interests are
\begin{eqnarray}
&&\epem\to W^+W^-Z,\quad 
{(\ell_{4,5,6,7})}
\label{wwz}\\
&&\epem\to ZZZ.\quad \quad\quad
{(\ell_{4,5,6,7,10})}
\label{zzz}
\end{eqnarray} 
which are more useful at lower energies near the three-vector
boson threshold. At higher energies, the fusion processes 
$W^*_LW^*_L \to \wlwl$ become dominant
\begin{eqnarray}
\label{fuse1}
&&\epem\to \bar\nu\nu W^+W^-,\quad 
{(\ell_{4,5}, H, \rho_T)} \\
&&\epem\to \bar\nu\nu ZZ,\quad\quad \quad
{(\ell_{4,5},\ell_{6,7}, H)}\\
&&\epem\to e^\pm \nu W^\mp Z,\qquad  
{(\ell_{4,5},\ell_{6,7}, \rho_T)}\\
&&\epem\to \epem ZZ.\quad \quad
(\ell_{4,5}+2\ell_{6,7,10})
\label{fuseL}
\end{eqnarray} 
With the linear collider running at different beam modes,
certain other processes can be complementary for the study
\begin{eqnarray}
\label{ememeg}
&&\emem\to \nu\nu W^-W^-,\quad\quad \quad (\ell_{4,5},I=2)\\
&&\gamma\gamma\to W^+W^-ZZ,\  W^+W^-W^+W^- ,\\
&&e \gamma \to e W^+W^-,\ eZ\gamma,\ eZZ.\quad
(\omega_T, A_{1})
\label{o1a1}
\end{eqnarray} 
If the top quark plays a role in the EWSB, then the direct
probe to the physics associated with this sector may be 
via the subprocess $\wlwl \to t\bar t$ for 
\begin{eqnarray}
\epem\to  \bar\nu\nu W^*W^* \to  \bar\nu\nu \bar t t.\quad \qquad
{(H, \rho_T...)}
\label{wt}
\end{eqnarray}
We now summarize the current results for studies of 
the above processes.

\subsection{Triple gauge-boson production}

\begin{figure}[th]
\psfig{figure=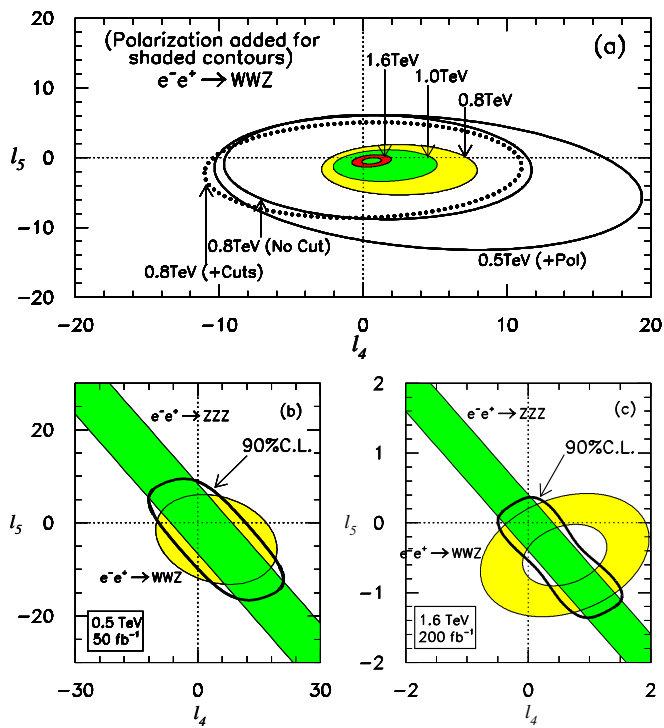,height=4.5in}
\vskip -1.5cm
\caption{$1\sigma$ contours in $\ell_4-\ell_5$ plane 
via $WWZ,ZZZ$ final states. We take $\Lambda=2$
TeV. The thick solid lines are for the two-channel combined 90\%
C.L. bounds.
\label{fig:vvv}}
\end{figure}

Triple gauge-boson production at $\epem$ linear colliders
will be an ideal process to study the quartic gauge
boson couplings.\cite{3v,vvv,ours} 
As outlined in Eqs.~(\ref{wwz})-(\ref{zzz}), 
the processes $\epem\to W^+W^-Z$ 
and $ZZZ$, involving $WWZZ,ZZZZ$ couplings, can be
sensitive to new physics. In Fig.~(\ref{fig:vvv}),
we present the sensitivity contours \cite{ours} at 90\%
C.L. for these two processes at different energies.
We see that with $\sqrt s=500$ GeV and an integrated
luminosity of 50 fb$^{-1}$, the magnitude of the 
couplings can be probed to about $4-10$. 
At $\sqrt s=1.6$ TeV with 200 fb$^{-1}$, the 
sensitivity to the couplings can be reached to
a level of $0.3-0.6$, reaching theoretically quite
interesting region. 
Beam polarizations of 90\% for $e^-$ and 65\% for $e^+$ 
helped to reduce the SM background.

\subsection{WW fusion processes}

\begin{figure}[tbh]
\psfig{figure=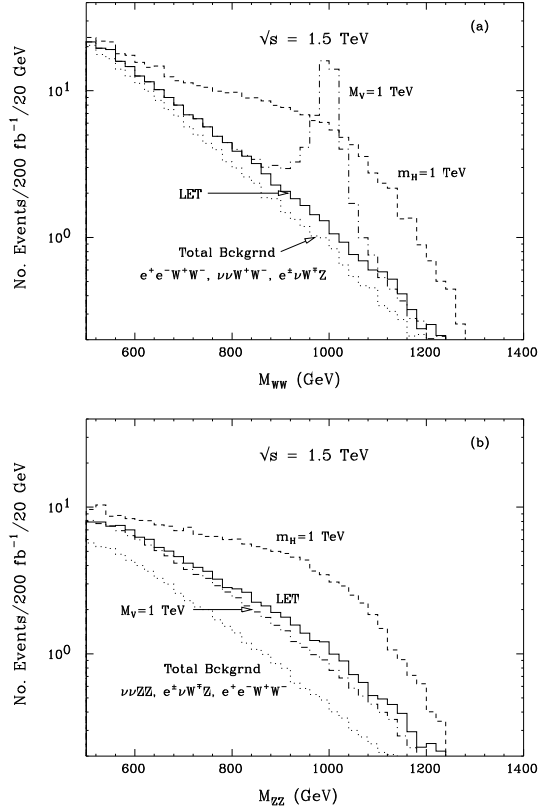,height=5.3in}
\vskip -1.8cm
\caption{Scalar and vector resonant production via 
(a) $W^+_L W^-_L \to W^+_LW^-_L$ and 
(b) $W^+_L W^-_L \to ZZ$ at a 1.5 TeV $e^+e^-$ linear 
collider. The summed SM background (dotted) and a LET 
amplitude (solid) are also presented.
\label{fig:bump}}
\end{figure}

Cross sections for the processes of Eqs.~(\ref{wwz})-(\ref{zzz}) 
are suppressed at higher energies,
typically like $s^{-1}$ well above the threshold. 
On the other hand, the gauge boson fusion processes
of Eqs.~(\ref{fuse1})-(\ref{fuseL}), originated by the
subprocesses 
$W^+W^-\to W^+W^-, ZZ; W^\pm Z\to W^\pm Z; ZZ \to ZZ$
and $W^-W^-\to W^-W^-$, become more 
significant, with an energy dependence typically like
$\ln(s/M_W^2)$ due to the collinear weak boson radiation 
off the electron beams in the initial state. In particular,
if there are new resonances in the SEWS sector that are
accessible at the collider, then the signal should be
more substantial. The backgrounds to the gauge-boson
fusion and the signal isolation technique 
have been extensively studied first at hadron
colliders \cite{wpwp,baggeretal} and later at the linear
colliders.\cite{wwbump} 
The SM backgrounds and SEWS signals
are calculated for a 1.5 TeV linear collider
for a scalar (Higgs-like), a vector
($\rho_{T}^{}$-like) of a mass one TeV, and the 
low-energy theorem amplitude (LET, non-resonance),
which are shown in Fig.~\ref{fig:bump}. 
The signal is clearly observable above the backgrounds
after judicious cuts.
It is important to note that by examining the
individual $W^+W^-$ or $ZZ$ final state, one may be
able to deduce the structure of the underlying dynamics.
For example, the ratio 
$\sigma(\wpwm\to\wpwm)/\sigma(\wpwm\to ZZ)$ is about two
for a Higgs-like scalar, is much larger than one for
a vector resonance, and is even smaller than one for 
the LET.\cite{wwbump}

\begin{figure}[tbh]
\psfig{figure=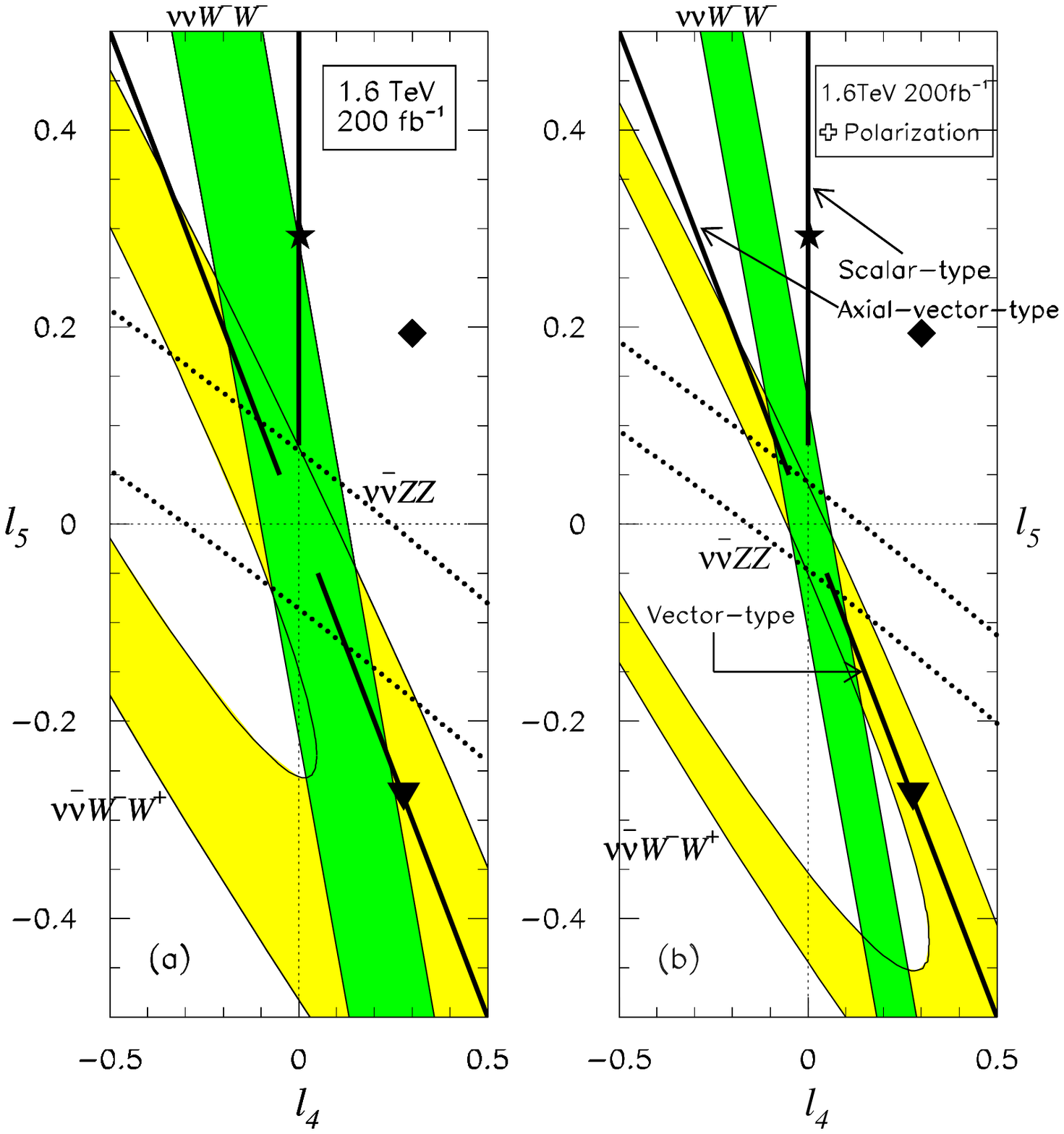,height=4.5in}
\vskip -1.0cm
\caption{$1\sigma$ contours in $\ell_4-\ell_5$ plane 
via $WW$ fusion processes with $W^+W^-,ZZ$ and $W^-W^-$ 
final states. We take $\Lambda=2$ TeV. 
\label{fig:vvvv}}
\end{figure}

For non-resonant scenario beyond the LET, the fusion processes 
are also studied \cite{wwfuse} for the couplings $\ell_{4,5,6,7,10}$, 
as shown in Fig.~\ref{fig:vvvv}. The characteristic range of 
the probe for a 1.6 TeV collider with 200 fb$^{-1}$ 
luminosity is about 0.06, which goes well below unity.
This sensitivity would be of great theoretical interest.

\subsection{$\wlwl\to t\bar t$}

\begin{figure}[tbh]
\psfig{figure=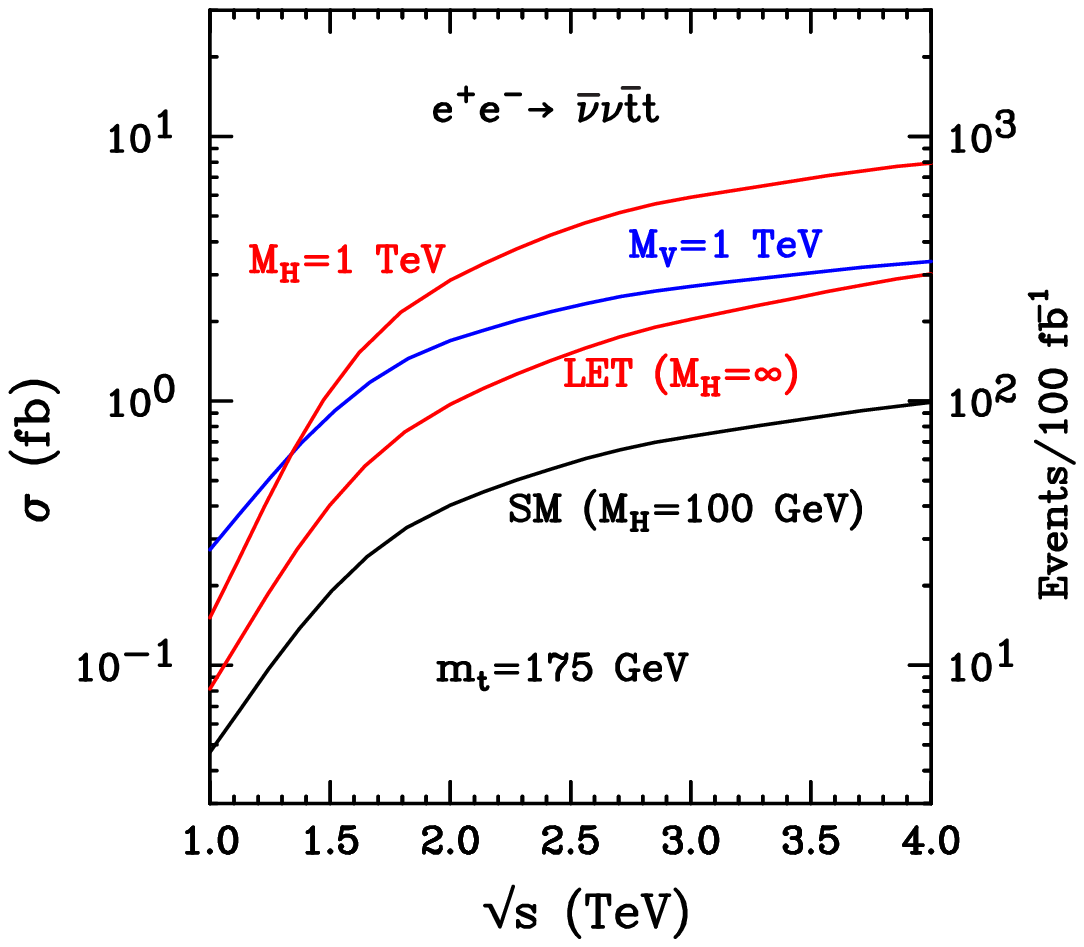,height=3.5in}
\vskip -0.5cm
\caption{Production cross section versus the c.m.~energy for 
$e^+e^-\to \bar \nu \nu \bar t t$.
The SM rate as the background is that of $m_H=100$ GeV;
the signal rates are labeled by $M_H=1$ TeV 
(a heavy scalar), $M_V=1$ TeV (a heavy vector)
and LET (leading order non-resonance).
\label{fig:tt}}
\end{figure}

The possibly significant role of the top quark played
in the EWSB sector motivates the study of the process 
Eq.~(\ref{wt}). The enhancement of the cross section 
due to the contribution from lower-lying resonances 
can be quite substantial and is shown in Fig.~\ref{fig:tt}
versus the $\epem$ c.m.~energy.\cite{wwtt2} We view
the SM result (with $m_H=100$ GeV) as the background. 
The signal rates are evaluated with the approximation of 
the Goldstone-boson equivalence theorem, and
are labeled by $M_H=1$ TeV (a heavy scalar), 
$M_V=1$ TeV (a heavy vector) and LET (leading order 
non-resonance). Another study on this channel was 
reported during the workshop \cite{wwtt1} 
and good signals in the $M_{t\bar t}$ distributions
for a scalar/vector resonances could be found. The
statistical significance for the signal channels
considered can be well above 10.

\section{Concluding remarks}\label{sec:conclude}

A few remarks are in order. First of all,
regarding the machines. We see that a TeV $e^+e^-$
linear collider should have great potential for
probing EWSB physics. On the other hand, 
$e^-e^-, e^-\gamma$ and $\gamma\gamma$ colliders
can be all complementary. As indicated in 
Eqs.~(\ref{ememeg})-(\ref{o1a1}),
the $\emem$ collision is unique in probing 
the weak isospin $I=2$ $W^-W^-$ scattering;\cite{emem} 
$e\gamma$ is unique in producing the weak isosinglet
states such as $\omega_T^{}$ via $\gamma Z$ 
fusion;\cite{egamma} and a $\gamma\gamma$ collider has 
similar potential to the $\epem$ collider.\cite{jikia}
Secondly, about the detectors. To effectively distinguish
a $W$ from a $Z$ in the hadronic modes, one would need
an adequate hadronic mass resolution for the calorimeter.
The detector coverage in the forward region should be
at least of the order of 10 degrees in order to tag/veto 
the forward leptons for the fusion processes.
Not mentioned in the presentation is the situation
where a vector resonance mixes with $\gamma/Z$ and the
mass is close or below the energy threshold. In this case, 
the signal would be particularly strong \cite{barklow}
and one can study its properties to a great detail.

Finally, we provide a ``naive'' comparison between a 
1.5 TeV linear collider  \cite{wwbump,ours,wwfuse} 
with a luminosity of 200 fb$^{-1}$ 
and the LHC \cite{baggeretal,lhcl} at 14 TeV with 
100 fb$^{-1}$. We see from Table~\ref{tabl}
that they have comparable reach for most of cases
under discussion. Due to the advantage of a cleaner 
experimental environment at the linear collider, 
the $t\bar t$ channel seems to be more accessible 
there than at the LHC. 

\begin{table}[th]
\caption{Comparison for the sensitivity for SEWS physics
between a 1.5 TeV $\epem$ collider and the LHC. The
entries under scalar, vector and $t\bar t$ are for the 
estimated statistical significance; those under
$\ell_{4,5}$ are the values that can be probed 
for the anomalous couplings.
.\label{tabl}}
\vspace{0.4cm}
\begin{center}
\begin{tabular}{|c|c|c|c|c|}\hline
 & scalar(1 TeV) & vector(1 TeV) & $\ell_{4,5}$& $t\bar t$ \\ \hline
LC(1.5 TeV) & 10 & 16 
            & $-0.1-$0.1 &     10   \\ \hline
LHC(14 TeV) & 8 & 15
            & $-0.5-$0.9 &not established \\ \hline
\end{tabular}
\end{center}
\end{table}

\section*{Acknowledgments}
I would like to thank the organizers of the workshop
for their invitation.
I also thank Hong-Jian He and German Valencia
for discussions during the preparation of this talk.
This work was supported in part by a DOE grant No.~DE-FG02-95ER40896 
and in part by the Wisconsin Alumni Research Foundation.

\end{document}